\documentclass[aps,preprint,amsmath,amssymb]{revtex4}
\usepackage{graphics,epsfig}
\newcommand{\bd}{\begin{document}}
\newcommand{\ed}{\end{document}}
\newcommand{\bc}{\begin{center}}
\newcommand{\ec}{\end{center}}
\newcommand{\be}{\begin{eqnarray}}
\newcommand{\ee}{\end{eqnarray}}
\renewcommand{\thefootnote}{\alph{footnote}}
\renewcommand{\theenumi}{\roman{enumi}}
\newcommand{\se}{\section}
\newcommand{\sse}{\subsection}
\newcommand{\bi}{\bibitem}

\newcommand{\func}{\rm}

\begin{document}

\begin{titlepage}
\begin{flushright}
\today
\end{flushright}
 \vskip 0.05in
 \null
\begin{center}
 \vspace{.15in}
{\Large {\bf
 T violation in $\Lambda_{b}\rightarrow \Lambda  \ell^+
\ell^-$ decays with polarized $\Lambda$ }
}\\
\vspace{1.0cm}  \par
 \vskip 2.1em
 {\large
  \begin{tabular}[t]{c}
{\bf Chuan-Hung Chen$^a$, C.~Q.~Geng$^{b,c}$ and J.~N.~Ng$^c$}
\\
\\
       {\sl ${}^a$Institute of Physics, Academia Sinica}
\\   {\sl  $\ $  Taipei, Taiwan 115, Republic of China }
\\
{\sl ${}^b$Department of Physics, National Tsing Hua University}
\\  {\sl  $\ $ Hsinchu, Taiwan, Republic of China }
\\
{\sl ${}^c$Theory Group, TRIUMF, 4004 Wesbrrok Mall}
\\ {\sl $\ $ Vancouver, B.C. V6T
2A3, Canada}
   \end{tabular}}
 \par \vskip 5.3em

 {\Large\bf Abstract}
\end{center}

We study the T violating effects in the baryonic decays of
$\Lambda_b\to\Lambda l^+l^-\ (l=e\,,\mu)$ with polarized
$\Lambda$.
We show that the transverse $\Lambda$ polarizations in these
baryonic decays could be as large as $50\%$ in CP violating
theories beyond the standard model such as the SUSY models,
which can be tested in various future hadron colliders.


\end{titlepage}


Recently, time-reversal violation (TV) has been measured
experimentally in the $K^0$ system \cite{TVk}, and thus
complements the information on CP violation (CPV) that has been
steadily accumulating for the past thirty seven years. The data is
in accordance with the CPT theorem which is fundamental to local
quantum field theories with Lorentz invariance and the usual
spin-statistics connection. However, the origin of the violation
remains unclear. In the standard model, CPV or TV arises from a
unique physical phase in the Cabibbo-Kobayashi-Maskawa (CKM) quark
mixing matrix \cite{ckm}. This paradigm also predicts CPV effect
in the b-quark system. To test the accuracy of this paradigm and
to search for other sources of CPV one needs to look for new
processes, especially the $B^0$ system. Indeed this is an
important quest of the B-factories. In addition we deem it
particularly interesting if the time reversal symmetry violation
can be directly detected in the b-system rather than inferring it
as a consequence of CPT invariance. In this paper, we will study
the T violating effects in the baryonic decays of
$\Lambda_b\to\Lambda l^+l^-$.

 It is known that for a general three-body decay of a baryon the triple 
spin-momentum
correlations, such as $\vec{s}\cdot \left( \vec{p}_{i}\times \vec{p}%
_{j}\right) $, are T-odd observables, where $\vec{s}$ and
$\vec{p}_{i,j}$ are the spin and momentum vectors of the final
particles, respectively. There are a number of different sources
that might give rise to these T-odd observables. The most
important ones being the weak CPV such as the CKM phase of the SM.
 However, final state
interactions such as QCD for non-leptonic decays
 or the electromagnetic (EM) interaction
among the final state particles can also make contributions. These
are usually less interesting and they could even hide the signals
from the weak CPV. We note that the T-odd triple correlations do
not need non-zero strong phases unlike some of CP violating
observables, such as the rate asymmetry between a particle and its
antiparticle.

In $\Lambda _{b}\rightarrow \Lambda l^{+}l^{-}$ decays, due to
experimental concerns only one polarization state can be measured. We can
 use either the polarization of the lepton ($s_{l}$) or the $\Lambda $ 
baryon ($s_{\Lambda}$)
to study the T-odd correlations.  For the case with a polarized lepton,
since the T-odd correlation such as the transverse lepton
polarization is always associated with the lepton mass, we expect
that the T violating effects are small for the light lepton modes
\cite{CQ}. Although the $\tau $ mode has less
suppression due to its mass, the corresponding branching ratio
which is $O(10^{-7})$ is about one order of magnitude smaller than
 that for the $e$ and $\mu $ ones. The above considerations plus  the fact
that the efficiency of
spin measurements in general are not
high make lepton polarization a poor choice. In our following analysis, we
will concentrate on the search for the
possibility of large T-odd term such as $\vec{s}_{\Lambda }\cdot (\vec{p}%
_{l^{+}}\times \vec{p}_{\Lambda })$
in  $\Lambda _{b}\rightarrow \Lambda l^{+}l^{-}$ with $l=e$ and $\mu $,
and we can set  $m_{l}=0$ within the accuracy of our calculations.

We start with the effective Hamiltonian for $b\rightarrow sl^{+}l^{-}$ by
including the right-handed couplings in the hadronic sector are given by

\begin{eqnarray}
{\cal H}\left( b\rightarrow sl^{+}l^{-}\right) &=&\frac{G_{F}\alpha _{em}}{%
\sqrt{2}\pi }V_{tb}V_{ts}^{*}\left[ H_{1\mu }L_{V}^{\mu }\ +H_{2\mu
}L_{A}^{\mu }\right]  \label{hameff}
\end{eqnarray}
with
\begin{eqnarray}
H_{1\mu } &=&\bar{s}\gamma ^{\mu }\left(
C_{9}^{L}P_{L}+C_{9}^{R}P_{R}\right) b -\frac{2m_{b}}{q^{2}}\bar{s}i\sigma
_{\mu \nu }q^{\nu }\left( C_{7}^{L}P_{R}+C_{7}^{R}P_{L}\right) b\,,
  \nonumber
\\
H_{2\mu } &=&\bar{s}\gamma ^{\mu }\left(
C_{10}^{L}P_{L}+C_{10}^{R}P_{R}\right) b \,, \nonumber \\
L_{V}^{\mu } &=&\bar{l}\gamma ^{\mu }l\,,  \nonumber \\
L_{A}^{\mu } &=&\bar{l}\gamma ^{\mu }\gamma _{5}l\,,
  \label{hameff1}
\end{eqnarray}
where $C_{i}^{L}$ and $C_{i}^{R}\left( i=7,9,10\right) $ denote the
effective Wilson coefficients of left- and right-handed couplings,
respectively. In the standard model,
\begin{eqnarray}
C_{9}^{L}=C_{9}^{eff}\,, &&C_{9}^{R}=0\,,  \nonumber \\
C_{10}^{L}=C_{10}\,, &&C_{10}^{R}=0\,,  \nonumber \\
C_{7}^{L}=C_{7}^{eff}\,, &&C_{7}^{R}={\frac{m_{s}}{m_{b}}}C_{7}^{eff}\,,
\label{SMWC}
\end{eqnarray}
where $C_{9}^{eff}$, $C_{10}$, and $C_{7}^{eff}$ are the standard Wilson
coefficients \cite{Buras}.

To study the exclusive decays of $\Lambda_b\to\Lambda l^+l^-$, one needs to
know the form factors in the transition of $\Lambda _{b}(p_{\Lambda
_{b}})\rightarrow \Lambda ( p_{\Lambda })$, parametrized generally as
follows:

\begin{eqnarray}
\left\langle \Lambda \right| \bar{s}\ \Gamma _{\mu }\ b\left| \Lambda
_{b}\right\rangle &=&f_{1}^{(T)}\bar{u}_{\Lambda }\gamma _{\mu }u_{\Lambda
_{b}}+f_{2}^{(T)}\bar{u}_{\Lambda }i\sigma _{\mu \nu }\ q^{\nu }u_{\Lambda
_{b}}+f_{3}^{(T)}q_{\mu }\bar{u}_{\Lambda }u_{\Lambda _{b}},  \nonumber \\
\left\langle \Lambda \right| \bar{s}\ \Gamma _{\mu }\gamma _{5}\ b\left|
\Lambda _{b}\right\rangle &=&g_{1}^{(T)}\bar{u}_{\Lambda }\gamma _{\mu
}\gamma _{5}u_{\Lambda _{b}}+g_{2}^{(T)}\bar{u}_{\Lambda }i\sigma _{\mu \nu
}\ q^{\nu }\gamma _{5}u_{\Lambda _{b}}+g_{3}^{(T)}q_{\mu }\bar{u}_{\Lambda
}\gamma _{5}u_{\Lambda _{b}}\,,  \label{atcq}
\end{eqnarray}
where $\Gamma_{\mu}=\gamma _{\mu }\ (i\sigma_{\mu\nu})$,
and $f_i^{(T)}$ and $g_i^{(T)}$ are the form factors of vector (tensor) and
axial-vector (axial-tensor) currents, respectively. In the heavy quark
effective theory (HQET) \cite{MR}, the form factors in Eq. (\ref{atcq}) can
be simplified by using
\begin{eqnarray}
\left\langle \Lambda (p_{\Lambda })\right| \bar{s}\Gamma b\left| \Lambda
_{b}(p_{\Lambda _{b}})\right\rangle =\bar{u}_{\Lambda }\left( F_{1}(q^{2})+%
\slash{\!\!\!{v}}F_{2}(q^{2})\right) \Gamma u_{\Lambda _{b}}\,,
\end{eqnarray}
where $\Gamma $ denotes the Dirac matrix, $v=p_{\Lambda _{b}}/M_{\Lambda
_{b}}$ is the four-velocity of $\Lambda _{b}$, and $q=p_{\Lambda
_{b}}-p_{\Lambda }$ is the momentum transfer, and the relations among the
form factors can be found in Ref. \cite{chen-prd63}. Explicitly, under the
HEQT, we have
\begin{eqnarray}
f_{1} &=& g_{2} = f^{T}_{2}=g^{T}_{2} = F_{1}+\sqrt{r}F_{2}\,,
  \nonumber 
\\
\rho &\equiv &M_{\Lambda _{b}}\left( {\frac{f_{2}+g_{2}}{f_{1}+g_{1}}}%
\right) = \frac{M_{\Lambda _{b}}}{q^2}\left( {\frac{f^{T}_{1}+g^{T}_{1}} 
{f_{1}+g_{1}}}%
\right) = {\frac{F_{2}}{F_{1}+\sqrt{r}F_{2}}}\,.  \label{HQETff}
\end{eqnarray}

In order to study the T violating effects using  the $\Lambda $ spin
polarization, we write the $\Lambda $ four-spin vector in terms of a unit
vector, $\hat{\xi}$, along the $\Lambda $ spin in its rest frame, as
\begin{eqnarray}
s_{0}\,=\,\frac{\vec{p}_{\Lambda }\cdot \hat{\xi}}{M_{\Lambda }},\qquad \vec{
s}\,=\,\hat{\xi}+\frac{s_{0}}{E_{\Lambda }+M_{\Lambda }}\vec{p}_{\Lambda },
\end{eqnarray}
and choose the unit vectors along the longitudinal, normal, transverse
components of the $\Lambda $ polarization, to be
\begin{eqnarray}
\hat{e}_{L} &=&\frac{\vec{p}_{\Lambda }}{\left| \vec{p}_{\Lambda }\right| },
\nonumber \\
\hat{e}_{N} &=&\frac{\vec{p}_{\Lambda }\times \left( \vec{p}_{l^{-}}\times
\vec{p}_{\Lambda }\right) }{\left| \vec{p}_{\Lambda }\times \left( \vec{p}%
_{l^{-}}\times \vec{p}_{\Lambda }\right) \right| },  \nonumber \\
\hat{e}_{T} &=&\frac{\vec{p}_{l^{-}}\times \vec{p}_{\Lambda }}{\left| \vec{p}%
_{l^{-}}\times \vec{p}_{\Lambda }\right| }\,,  \label{uv}
\end{eqnarray}
respectively. Hence, the differential decay rates with polarized
$\Lambda $ is given by
\begin{eqnarray}
d\Gamma &=&\frac{1}{2}d\Gamma ^{0}\left[ 1+\vec{P}\cdot 
\hat{\xi}\right]\,,
\label{diffrate} \\
d\Gamma ^{0}\left( t\right) &=&\frac{G_{F}^{2}\alpha _{em}^{2}\lambda
_{t}^{2}}{96\pi ^{5}}M_{\Lambda _{b}}^{5}\sqrt{t^{2}-r}f_{1}^{2}R_{\Lambda
_{b}}\left( t\right) dt,
\label{diffrate0}
\end{eqnarray}
where $\vec{P}$ is the $\Lambda$ polarization vector, defined by
\begin{eqnarray}
\vec{P}=P_{L}\hat{e}_{L}+P_{N}\hat{e}_{N}+P_{T}\hat{e}_{T}\,,
\end{eqnarray}
and
\begin{eqnarray}
R_{\Lambda _{b}}\left( t\right) &=&\left[ \left( 3\left( 1+r\right)
t-2r-4t^{2}\right) +s\rho ^{2}\left( 3\left( 1+r\right) t-4r-2t^{2}\right)
+6\rho \sqrt{r}s\left( 1-t\right) \right]  \nonumber \\
&&\times \left( \left| C_{9}^{R}\right| ^{2}+\left| C_{9}^{L}\right|
^{2}+\left| C_{10}^{R}\right| ^{2}+\left| C_{10}^{L}\right| ^{2}\right)
\nonumber \\
&&+\frac{4\hat{m}_{b}}{s}\left( 3\left( 1+r\right) t-4r-2t^{2}\right) \left(
\left| C_{7}^{R}\right| ^{2}+\left| C_{7}^{L}\right| ^{2}\right)  \nonumber
\\
&&-6\sqrt{r}s\left[ 1+2\left( t-r\right) \rho +s\rho ^{2}\right] \left( {\rm
{Re}C_{9}^{R}C_{9}^{L*}+{Re}C_{10}^{R}C_{10}^{L*}}\right)  \nonumber \\
&&+12\hat{m}_{b}\left[ 2\rho \left( r+t^{2}-\left( 1+r\right) t\right) -%
\sqrt{r}\left( 1-t\right) \left( 1+s\rho ^{2}\right) \right] \left( {\rm {Re}%
C_{9}^{R}C_{7}^{L*}+{Re}C_{9}^{L}C_{7}^{R*}}\right)  \nonumber \\
&&+12\hat{m}_{b}\left[ 2\rho \sqrt{r}s+\left( t-r\right) \left( 1+s\rho
^{2}\right) \right] \left( {\rm {Re}C_{9}^{R}C_{7}^{R*}+{Re}%
C_{9}^{L}C_{7}^{L*}}\right) \,,
\end{eqnarray}
with $\lambda _{t}=V_{tb}V_{ts}^{*}$, $t=E_{\Lambda }/M_{\Lambda _{b}}$, $%
r=M_{\Lambda }^{2}/M_{\Lambda _{b}}^{2}$, $\hat{m}_{b}=m_{b}/M_{\Lambda
_{b}}$, and $s=1+r-2t$. The kinematic ranges for $t$ and $s$ are
\begin{eqnarray}
\sqrt{r}\leq t &\leq &\frac{1}{2}\left( 1+r\right) \,, \nonumber
\\
0\leq s &\leq &\left( 1-\sqrt{r}\right) ^{2}\,.
\end{eqnarray}
 With the T odd transverse $\Lambda $ polarizations defined by
\[
P_{T}=\frac{d\Gamma \left( \hat{\xi}\cdot \hat{e}_{T}=1\right) -d\Gamma
\left( \hat{\xi}\cdot \hat{e}_{T}=-1\right) }{d\Gamma \left( \hat{\xi}\cdot
\hat{e}_{T}=1\right) +d\Gamma \left( \hat{\xi}\cdot \hat{e}_{T}=-1\right) }
\,,
\]
 and from Eq. (\ref{diffrate}) we obtain
\begin{eqnarray}
P_{T} &=&\frac{3\pi }{4R_{\Lambda _{b}}\left( t\right) }\sqrt{s\phi }\left(
1-s\rho ^{2}\right)  \nonumber \\
&&\times \left[ \left( {\rm {Im}C_{9}^{R}C_{10}^{L*}-{Im}C_{9}^{L}C_{10}^{R*}%
}\right) -\frac{2\hat{m}_{b}}{s}\left( 1-t\right) \left( {\rm {Im}%
C_{7}^{L}C_{10}^{R*}-{Im}C_{7}^{R}C_{10}^{L*}}\right) \right] \;\;,  \label{Pt}
\end{eqnarray}
where $\phi \left( s\right) =\left( 1-r\right) ^{2}-2s\left( 1+r\right)
+s^{2}$.

To understand Eq. (\ref{Pt}), we first examine the hadronic currents with $%
V\pm A$ types of interactions.  From Eq. (\ref{hameff}), there are three
possible sources  which
could lead to $T$-odd correlations:
\renewcommand{\labelenumi}{(\theenumi)}
\begin{enumerate}
\item $H_{i\mu}H_{i\nu}^{\dagger} 
L_{V(A)}^{\mu}L_{V(A)}^{\nu\dagger}\,,\;\;\;\;\;\ (i=1,2)$
\item  $H_{1\mu }^{L(R)}H_{2\nu
}^{L(R)\dagger }L_{V}^{\mu }L_{A}^{\nu \dagger}$,
\item  $H_{1\mu
}^{L(R)}H_{2\nu }^{R(L)\dagger }L_{V}^{\mu }L_{A}^{\nu\dagger }$,
\end{enumerate}
 where $H_{i\mu }^{L(R)}$ involve only
left-handed (right-handed) currents. By summing all the lepton spin degrees
of freedom, we get $\sum L_V^{\mu}L_V^{\nu\dagger}=\sum L_A^{\mu}
L_A^{\nu\dagger}=(p_l^{\mu}p_{\bar{l}}^{\nu}+p_l^{\nu}p_{\bar{l}}^{\mu}
-g^{\mu\nu}p_l\cdot p_{\bar{l}})$, which are symmetric with respect to $\mu$
and $\nu$. Since $H_{i\mu}H_{i\nu}^{\dagger}\propto\varepsilon_{\mu\nu\alpha%
\beta} s_{\Lambda}^{\alpha}p_{\Lambda_b}^{\beta}$ is antisymmetric between $%
\mu$ and $\nu$, it is clear that no T-odd terms can be constructed from (i).
For the case in (ii), even though $ \sum L_V^{\mu}L_A^{\nu\dagger}
=-4i\varepsilon^{\mu\nu\alpha\beta}p_{l\alpha}p_{\bar{l}\beta}$ is
antisymmetric, 
$ H_{1\mu}^{L(R)}H_{2\nu}^{L(R)\dagger}\propto$ \\
$ M_{\Lambda}Tr(\slash{\!\!\!s}_{\Lambda}\gamma_{\mu}
\slash{\!\!\!p}_{\Lambda_b}
\gamma_{\nu})=4M_{\Lambda}\left(s_{\Lambda\mu}p_{\Lambda_b\nu }
+s_{\Lambda\nu}p_{\Lambda_b\mu}-g_{\mu\nu}s_{\Lambda} \cdot
p_{\Lambda_b}\right)$ is symmetric in $\mu $ and $\nu$ and thus, the
possible terms in (ii) also vanish. However, T-odd correlations can arise
from (iii) by observing that $H_{1\mu}^{L(R)}H_{2\nu}^{R(L)\dagger}\propto
M_{\Lambda_b}Tr (\not{\!}\!{p}_{\Lambda}\not{\!}\!{s}_{\Lambda}\gamma_{\mu}%
\gamma_{\nu})
=4M_{\Lambda_b}\left(-p_{\Lambda\mu}s_{\Lambda\nu}+p_{\Lambda\nu}
s_{\Lambda\mu}\right)$ is antisymmetric.  Hence, in $\Lambda _{b}\rightarrow
\Lambda l^{+}l^{-}$ decays, the non-vanished T-odd terms can be induced from
 $\left( V-A\right) \times \left( V+A\right)$ hadronic currents. This is, 
explicitly
 shown in Eq. (\ref{Pt}). The TV quantity is related to 
Im$C_{9}^{R}C_{10}^{L*}$ and Im%
$C_{9}^{L}C_{10}^{R*}$, respectively. Similarly, with the dipole operators,
we expect that T-odd observables are proportional to Im$C_{7}^{R}C_{10}^{L*}$
and Im$C_{7}^{L}C_{10}^{R*}$.

As seen in Eq. (\ref{Pt}), to have a non-zero value of $P_T$, it
is necessary to have conditions of (i) the existence of $C_k^R$
and (ii) a phase of $C_i^LC_j^R\ (i\neq j)$. These conditions are
clearly different from those of the T odd transverse lepton
polarizations in both inclusive decay of $b\to s l^+l^-$
\cite{Tin} and exclusive ones, such as $B\to K^{(*)}l^+l^-$
\cite{TexM} and $\Lambda_b\to\Lambda l^+l^-$ \cite{TexB}, and
other CP violating asymmetries as well \cite{CQ,HK}.
Such property
distinguishes the T odd transverse $\Lambda$ polarizations from various
other T odd observables. In the standard model, since there are no $C^R_{9}$
and $C^R_{10}$ as seen from Eq. (\ref{SMWC}),
\begin{eqnarray}
P_T^{SM}&\propto& {\frac{m_s}{m_b}}Im\left(C_7^{eff}C_{10}\right)\,,
\end{eqnarray}
which is suppressed. We note that the contribution to $P_T$
from the EM final state interaction is $<O(10^{-3})$. Moreover, the long-distance (LD)
effects in the one-loop matrix elements of
$O_{1,2}$ \cite{Buras} and $\Lambda_b \rightarrow \Lambda J/\Psi$
with $J/\Psi\rightarrow l^+ l^-$
are absorbed to $C^{L}_{9}$. On the
other hand, it is clear that a large value of $P_T$, according to
Eq. (\ref{Pt}), can be obtained if a theory contains $C_9^R$ or
$C_{10}^R$ or a large $C_7^R$, with a non-zero phase.
Many theories beyond the
standard model could give rise to $C_9^R$ or $C_{10}^R$; examples are
the left-right symmetric and supersymmetric models.

To illustrate our result, we use
SUSY models both with
and without R-parity. For a SUSY model with R-parity, we take the one
given
by Ref. \cite{Masiero}, denoted as $M1$, in which
\begin{eqnarray}
C_{7}^{L\,M1} &=&-1.75\left( \delta _{23}^{u}\right) _{LL}-0.25\left( \delta
_{23}^{u}\right) _{LR}-10.3\left( \delta _{23}^{d}\right) _{LR} \,,
\nonumber \\
C_{9}^{L\,M1} &=&0.82\left( \delta _{23}^{u}\right) _{LR} \,,  \nonumber \\
C_{10}^{L\,M1} &=&-9.37\left( \delta _{23}^{u}\right) _{LR}+1.4\left( \delta
_{23}^{u}\right) _{LR}\left( \delta _{33}^{u}\right) _{RL}+2.7\left( \delta
_{23}^{u}\right) _{LL}\,,  \nonumber \\
C_{7}^{R\,M1} &=&-10.3\left( \delta _{23}^{d}\right) _{RL} \,,  \nonumber \\
C_{9}^{R\,M1} &=&1.32\left( \delta _{23}^{d}\right) _{RL}\left( \delta
_{33}^{d}\right) _{LR}\,,  \nonumber \\
C_{10}^{R\,M1} &=&-17.6\left( \delta _{23}^{d}\right) _{RL}\left( \delta
_{33}^{d}\right) _{LR}\,,  \label{susywr}
\end{eqnarray}
where the values of parameters $\delta_{ij}^q$ are as follows :
\begin{eqnarray}
\left( \delta _{23}^{d}\right) _{LR} & \sim & \left( \delta _{23}^{d}\right)
_{RL} \:\sim \;-3\times 10^{-2}e^{i\theta _{1}}\,,  \nonumber \\
\left( \delta _{33}^{d}\right) _{LR} & \sim & \left( \delta _{33}^{u}\right)
_{RL}\: \sim \: 0.5\,,  \nonumber \\
\left( \delta _{23}^{u}\right) _{LR} & \sim & -0.7e^{i\theta _{2}}\,,  
\nonumber
\\
\left( \delta _{23}^{u}\right) _{LL} & \sim & 0.1\,,
\end{eqnarray}
with taking the phases of $\theta_1$ and $\theta_2$ to be $\pi /4$ and $-\pi
/10$, respectively.

 To demonstrate the difference we study a  SUSY model without R-parity
($M2$, see Ref. \cite{Nardi}). In this case we have

\begin{eqnarray}
C_{9}^{R\,M2} &=&-C_{10}^{R\,M2}\;=\;\frac{\pi }{\sqrt{2}G_{F}\alpha _{em}}%
\frac{1}{V_{tb}V_{ts}^{*}}\frac{\lambda _{ij3}^{\prime *}\lambda
_{ij2}^{\prime }}{2M_{\tilde{u}_{j}}^{2}}  \label{susynr}
\end{eqnarray}
where
\begin{eqnarray}
\left| \lambda _{ijk}^{\prime }\right| &\simeq &\left( 2\sqrt{2}G_{F}\tan
^{2}\beta \right) ^{1/2}\zeta _{i}m_{d_{j}}\lambda^{2}  \label{susynr1}
\end{eqnarray}
and
\begin{eqnarray}
\lambda _{ij3}^{\prime *}\lambda _{ij2}^{\prime } &\simeq &2\sqrt{2}%
G_{F}\tan ^{2}\beta \zeta _{i}^{2}m_{d_{j}}^{2}\lambda^{4}e^{i\theta _{\not%
{R}}}  \label{susynr2}
\end{eqnarray}
with the parameters $\lambda \sim 0.22$, $\tan \beta \sim 10$, $\zeta _{i}
\sim 0.9$, and $\theta _{\not{R}} =2\pi/5$.
We note that the parameters we have chosen for 
the two models satisfy various experimental
constraints \cite{Masiero,Nardi}.

\begin{figure}[htbp]
 \centerline{\psfig{figure=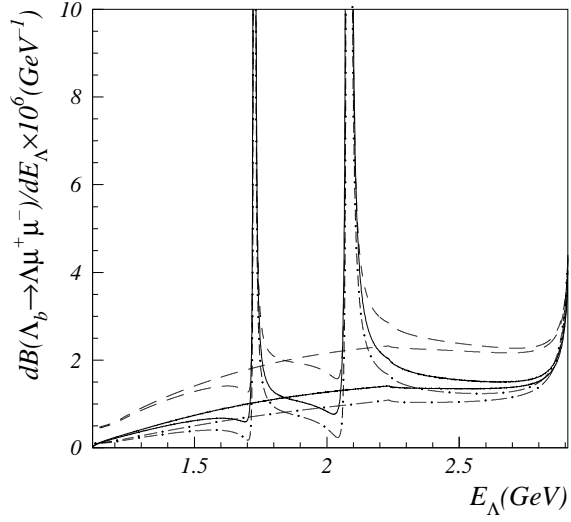,height=2.7in }
}\caption{ Differential BRs of $\Lambda_b\to\Lambda \mu^+\mu^-$ as
a function of $E_{\Lambda}$ with and without resonant shapes,
where the solid, dash-dotted and dashed curves stand for the
results of the standard and SUSY with and without R-parity models,
respectively. } \label{rpmurate}
\end{figure}

By using the parameters in Eqs. (\ref{susywr})-(\ref{susynr2}), the values
for the form factors in Refs. \cite{CQ,chen-prd63},
and Eq. (\ref{diffrate0}),
we give the differential branching ratios (BRs) of
$\Lambda_b\to\Lambda\mu^+ \mu^-$ with respect to $E_{\Lambda}$
in Figure \ref{rpmurate} with and without including
resonant states of $\Psi$ and $\Psi ^{\prime }$,
and we find that
the integrated BRs for the latter are $(2.10,\,1.66,\,4.41)\times 10^{-6}$
for the standard and SUSY with
and without R-parity models, respectively.

\begin{figure}[htbp]
 \centerline{\psfig{figure=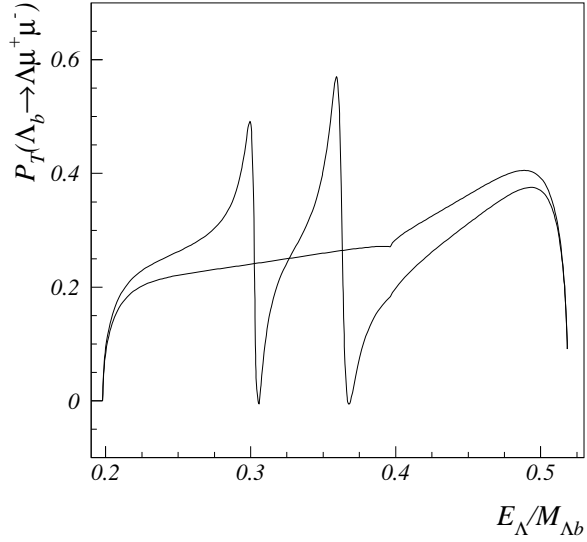,height=2.8in }}
\caption{ Transverse $\Lambda$ polarization in $\Lambda_b\to\Lambda
\mu^+\mu^-$ as a function of $E_{\Lambda}/M_{\Lambda_b}$ in the SUSY model
with R-parity. The curves with and without resonant shapes represent
including and no long-distance contributions, respectively. }
\label{susypt}
\end{figure}

\begin{figure}[htbp]
\centerline{ \psfig{figure=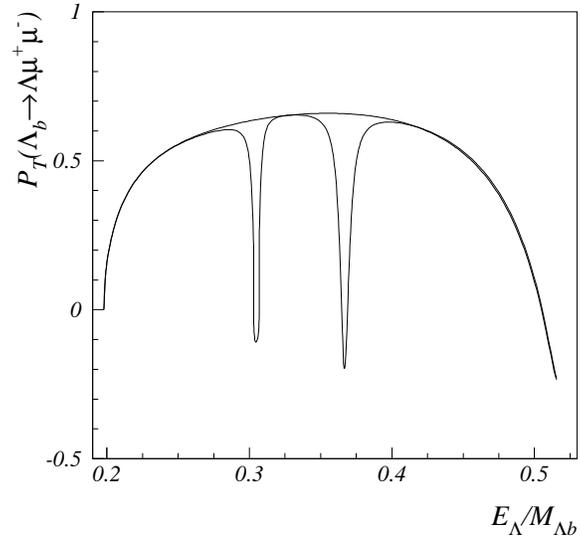,height=2.8in}}
\caption{ Same as Figure \ref{susypt}
but for the SUSY model without R-parity. } \label{rppt}
\end{figure}

In Figures \ref{susypt} and \ref{rppt}, we show
$P_T(\Lambda_b\to\Lambda \mu^+\mu^-)$ as a function of $%
E_{\Lambda}/M_{\Lambda_b}$
for $M1$ and $M2$, with and without the LD contributions,
respectively. As seen
from the figures, even though the derivations of BRs to the standard
model result are insignificant, the transverse $\Lambda$ polarization
asymmetries can be over $50\%$ in both SUSY models with and without
R-parity. Similar results are also expected the decay of
$\Lambda_b\to\Lambda e^+e^-$.
We remark that measuring a large $P_T$ in $\Lambda_b\to\Lambda
l^+l^-$ is a clean indication of T violation as well as new CP
violation mechanism beyond the standard model.

 Finally, we note
that to measure $P_T(\Lambda_b\to\Lambda
\mu^+\mu^-)\sim 10\%$ at $3\sigma$ level, at least $4.5\times 10^7$ $%
\Lambda_b$ decays are required if we use $BR(\Lambda_b\to\Lambda
\mu^+\mu^-)\sim 2\times 10^{-6}$.
Clearly, the measurement could be done in the second generation
of B-physics experiments, such as LHCb, ATLAS, and CMS at the LHC, and
BTeV at
the Tevatron, which produce $\sim 10^{12}b\bar{b}$ pairs per year \cite{BB}.
This is certainly within reach of a super B factory under discussion now
\cite{SuperB}.

\noindent {\bf Acknowledgments}

This work was supported in part by
 the National Center for Theoretical Science,
National Science Council of the Republic of China under
 Contract Nos. NSC-90-2112-M-001-069 and NSC-90-2112-M-007-040,
 and National Science and Engineering Research Council of Canada.


\newpage


\end{document}